\documentclass[a4paper,12pt]{article}
\usepackage[utf8x]{inputenc}
\usepackage{amsmath}
\usepackage{amssymb}
\usepackage{graphicx}
\usepackage{epstopdf}

\title{Solar and Atmospheric Neutrinos: Limitations for Direct Dark Matter Searches}
\author{A.~G\"utlein, W.~Potzel, C.~Ciemniak, F.~von~Feilitzsch,\\N.~Haag, M.~Hofmann, C.~Isaila, T.~Lachenmaier,\\J.-C.~Lanfranchi, L.~Oberauer, S.~Pfister, S.~Roth,\\M.~von~Sivers, R.~Strau\ss{}, and A.~Z\"oller\\\\Physik-Department E15,\\ Technische Universit\"at M\"unchen, D-85748 Garching,\\Germany}

\date{\today}

\begin{document}
\maketitle

\begin{abstract}
In experiments for direct dark matter searches, neutrinos coherently scattering off nuclei can produce similar events as Weakly Interacting Massive Particles (WIMPs). To reach sensitivities better than $\sim10^{-10}$\,pb for the elastic WIMP nucleon spin-independent cross section in the zero-background limit, energy thresholds for nuclear recoils should be $\gtrsim2.05$\,keV for CaWO$_4$, $\gtrsim4.91$\,keV for Ge, $\gtrsim2.89$\,keV for Xe, $\gtrsim8.62$\,keV for Ar and $\gtrsim15.93$\,keV for Ne as target material. Atmospheric neutrinos limit the achievable sensitivity for the background-free direct dark matter search to $\gtrsim10^{-12}$\,pb.
\end{abstract}

Coherent Neutrino Nucleus Scattering (CNNS) is a neutral-current interaction (exchange of a virtual Z$^{0}$ boson) and thus independent of the neutrino flavour. In a CNNS event, momentum and recoil energy are transferred to the nucleus. For low transferred momenta where the Z$^{0}$-wavelength and the radius of the nucleus are about equal, the neutrino scatters off all nucleons coherently leading to an enhanced cross section \cite{Drukier84}.

The count rate $R_{th}$ in a detector due to CNNS above a threshold $E_{th}$ of the nuclear recoil energy is given by \cite{Guetlein10}
\begin{equation}
R_{th} =\int_{E_{th}}^{\infty} dE_{rec} \frac{dR\left(E_{rec}\right)}{dE_{rec}} = N_t \int_{E_{th}}^{\infty} dE_{rec}\int_{\sqrt{\frac{E_{rec}M}{2}}}^{\infty}dE_{\nu}\Phi(E_{\nu})\frac{d\sigma(E_{\nu}, E_{rec})}{dE_{rec}},
\end{equation}
where $M$ is the mass of the target nucleus, $E_{rec}$ its recoil energy; $N_{t}$ is the total number of target nuclei, $\Phi(E_{\nu})$ the neutrino flux at a neutrino energy $E_{\nu}$; $\frac{dR\left(E_{rec}\right)}{dE_{rec}}$ is the recoil spectrum and $\frac{d\sigma(E_{\nu}, E_{rec})}{dE_{rec}}$ is the differential scattering cross section which includes the Helm form factor \cite{Helm56}, \cite{Guetlein10}, see below.

Assuming an isothermal WIMP halo, the recoil spectra
\begin{figure}[htb]
	\centering
	\includegraphics[width=0.8\textwidth]{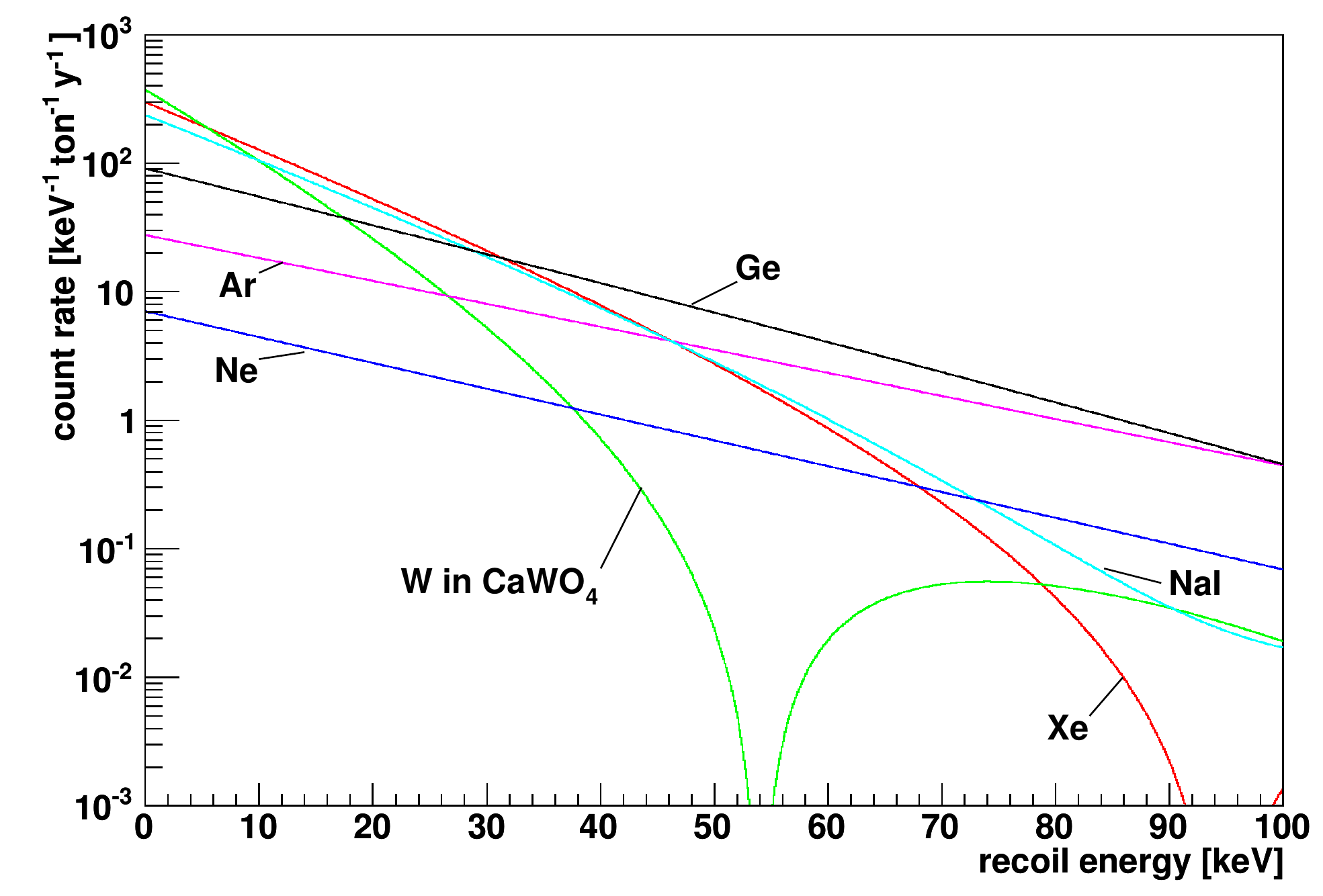}
	\caption{Recoil spectra of WIMPs for different target materials assuming a WIMP mass of 60\,GeV and a WIMP-nucleon cross section of $10^{-44}$\,cm$^2$.}
	\label{spectra}
\end{figure}
(see figure \ref{spectra}) of WIMPs scattering coherently off various target materials were calculated according to \cite{Lewin96}
\begin{equation}
	\frac{dR(E_{rec})}{dE_{rec}} = \frac{c_1 N_A \rho_D}{2\sqrt{\pi}\mu_1^2}\sigma_{WN}\left|F(q)\right|^2v_0\frac{A^2}{E_0}e^{\frac{-c_2 E_{rec}}{E_0 r}}
\end{equation}
where $c_1$, $c_2$ are constants describing the annular modulation of the WIMP flux, $N_A$ is Avogadro's number, $\rho_D$ the local WIMP density, $\mu_1$ the reduced mass for A=1, $\sigma_{WN}$ the WIMP-nucleon cross section, $F(q)$ is the Helm form factor \cite{Helm56}, $v_0$ the velocity of the earth relative to the galaxy, $A$ the mass number, $E_0$ the kinetic energy of the WIMPs; $r = 4\frac{M_D M_T}{(M_D + M_T)^2}$ a kinematic factor, $M_D$ the WIMP mass and $M_T$ the mass of the target nucleus.

Since recoil energies due to solar neutrinos are small, such events can be rejected by a proper choice of the energy threshold. If one allows only 0.1 counts per ton-year, the energy regions listed in the second column of
\begin{table}[htb]
	\centering
	\begin{tabular}{|l|r||r|r|}
		\hline
		\textbf{Material} & \textbf{energy region} & \textbf{max. exposure} & \textbf{energy region}\\
		& solar neutrinos &  & atmosph. neutrinos\\
		\hline
		Ne & 15.93 - 100\,keV & 35.9\,ton-years & 15.93 - 100\,keV\\
		\hline
		Ar & 8.62 - 100\,keV & 9.2\,ton-years & 8.62 - 100\,keV\\
		\hline
		Ge & 4.91 - 100\,keV & 4.3\,ton-years & 4.91 - 100\,keV\\
		\hline
		Xe & 2.89 - 100\,keV & 3.0\,ton-years & 2.89 - 50\,keV\\
		\hline
		NaI & 11.48 - 100\,keV & 6.5\,ton-years & 11.48 - 50\,keV\\
		\hline
		W in CaWO$_4$ & 2.05 - 100\,keV & 3.9\,ton-years & 2.05 - 40\,keV\\
		\hline
	\end{tabular}
	\caption{The second column denotes the optimal energy regions for the WIMP search taking solar neutrinos into account. In the third column the exposures leading to 0.1 atmospheric-neutrino events are given if an optimal energy region according to the last column is selected for the WIMP search.}
	\label{regions}
\end{table}
table \ref{regions} are obtained. As an upper limit, 100\,keV was assumed. This value is uncritical since the number of WIMP events drops exponentially with energy. The corresponding exclusion plots
\begin{figure}[h!]
	\centering
	\includegraphics[width=0.8\textwidth]{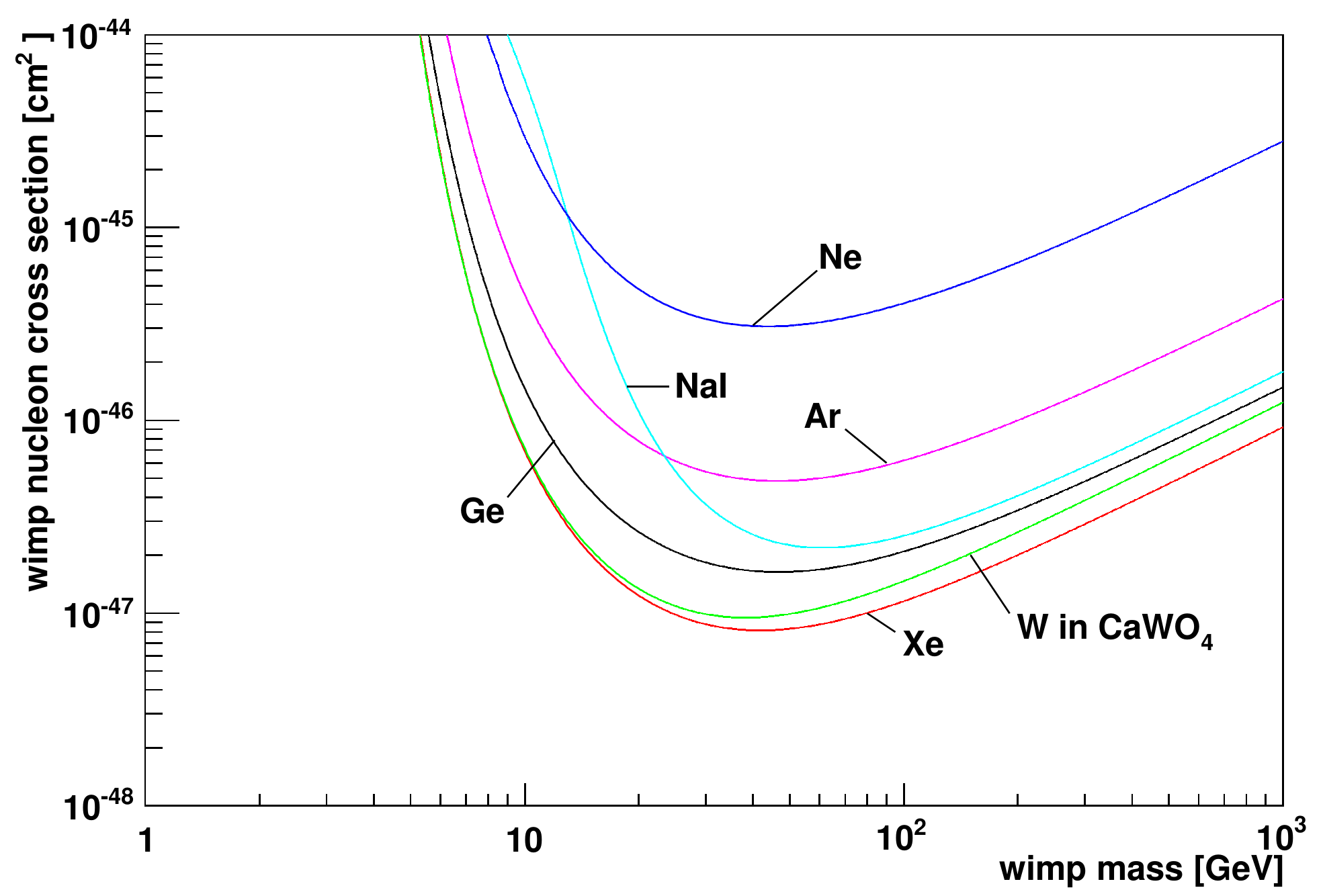}
	\caption{Exclusion plots for different target materials (Exposure 1\,ton-year, 90\% confidence level).}
	\label{exclusion}
\end{figure}
(see figure \ref{exclusion}) for various target materials were obtained for an exposure of 1 ton-year, 90\% confidence level assuming zero events in the optimal energy region of table \ref{regions} (second column).

For WIMP-nucleon cross sections below $10^{-46}$\,cm$^2$ ($10^{-10}$\,pb) solar neutrinos have to be taken into account. For WIMP masses $>10$\,GeV, Xe and W in CaWO$_4$ are the most promising target materials.

For sensitivities below $10^{-48}$\,cm$^2$ ($10^{-12}$\,pb) also atmospheric neutrinos are becoming an important background source \cite{Guetlein10}. The third column of table \ref{regions} shows the maximal exposure if only 0.1 atmospheric-neutrino events within the optimal energy window given by the last column are allowed. Again the upper limit of this energy window is found to be uncritical. For example, if for Xe the upper threshold is increased from 50 to 100\,keV, the maximum exposure is reduced to 2.9 ton-years.

This work was supported by funds of the Deutsche Forschungsgemeinschaft DFG (Transregio 27: Neutrinos and Beyond), the Munich Cluster of Excellence (Origin and Structure of the Universe), and the Maier-Leibnitz-Laboratorium (Garching).


\begin{thebibliography}{00}

\bibitem{Drukier84} A. Drukier and L. Stodolsky, Phys. Rev. D \textbf{30} (1984) 2295

\bibitem{Guetlein10} A. G\"utlein et al., Astropart. Phys. \textbf{34} (2010) 90

\bibitem{Helm56} R.H. Helm, Phys. Rev. \textbf{104} (1956) 1466

\bibitem{Lewin96} J.D. Lewin and P.F. Smith, Astropart. Phys. \textbf{6} (1996) 87

\end{thebibliography}
\end{document}